\documentclass[12pt]{article}
\input epsf.tex

\newdimen\SaveWidth \SaveWidth=\textwidth
\newdimen\SaveHeight \SaveHeight=\textheight
\advance\SaveWidth by -\textwidth
\advance\SaveHeight by -\textheight
\divide\SaveWidth by 2
\divide\SaveHeight by 2
\advance\hoffset by \SaveWidth
\advance\voffset by \SaveHeight
\def\abs#1{\left| #1\right|}

\def\etal{{\it et al.}}
\def\abs#1{\left| #1\right|}
\def\sgn{\mathop{\rm sgn}}
  
\def\etmiss{\slashchar{E}_T}

\def\tell{{\tilde\ell}}
\def\ttau{{\tilde\tau}}
\def\fbi{{\rm fb}^{-1}}
\def\Meff{M_{\rm eff}}

\def\lsp{{\tilde\chi_1^0}}

\def\GeV{{\rm GeV}}
\def\mhalf{m_{1/2}}
\def\tchi{\tilde\chi}
\def\tg{\tilde g}
\def\tq{\tilde q}

\let\badcite=\cite
\def\cite{~\badcite}

\def\cmsec{{\rm cm}^{-2}{\rm s}^{-1}}

\def\slashchar#1{\setbox0=\hbox{$#1$}           
   \dimen0=\wd0                                 
   \setbox1=\hbox{/} \dimen1=\wd1               
   \ifdim\dimen0>\dimen1                        
      \rlap{\hbox to \dimen0{\hfil/\hfil}}      
      #1                                        
   \else                                        
      \rlap{\hbox to \dimen1{\hfil$#1$\hfil}}   
      /                                         
   \fi}                                         %

\catcode`@=11
\newdimen\vbigd@men                             

\def\vbig#1#2{{\vbigd@men=#2\divide\vbigd@men by 2%
   \hbox{$\left#1\vbox to \vbigd@men{}\right.\n@space$}}}
\catcode`@=12

\def\simge{
    \mathrel{\rlap{\raise 0.511ex
        \hbox{$>$}}{\lower 0.511ex \hbox{$\sim$}}}}
\def\simle{
    \mathrel{\rlap{\raise 0.511ex 
        \hbox{$<$}}{\lower 0.511ex \hbox{$\sim$}}}}

\catcode`@=11
\def\citenum#1{\csname b@#1\endcsname}
\catcode`@=12

\def\dofig#1#2{\centerline{\epsfxsize=#1\epsfbox{#2}}}
\def\dofigs#1#2#3{\centerline{\epsfxsize=#1\epsfbox{#2}%
  \epsfxsize=#1\epsfbox{#3}}}

\begin{document}

\rightline{LBNL-45963}
\rightline{BNL-HET-00/17}
\rightline{ATL-COM-PHYS-2000-011}

\bigskip

\begin{center}{\Large\bf
Lepton Flavor Violation at the LHC\footnotemark}
\end{center}

\footnotetext{This work was supported in part by the Director,
Office of Science, Office of Basic Energy Research, Division of High
Energy Physics of the U.S. Department of Energy under Contracts
DE-AC03-76SF00098 and DE-AC02-98CH10886.}

\bigskip
\centerline{\bf I. Hinchliffe$^a$ and F.E. Paige$^b$
}
\centerline{$^a${\it Lawrence Berkeley National Laboratory, Berkeley, CA}}
\centerline{$^b${\it Brookhaven National Laboratory, Upton, NY}}
\bigskip

\begin{abstract}
        Recent results from Super Kamiokande suggest $\nu_\mu-\nu_\tau$
mixing and hence lepton flavor violation. In supersymmetric models, this
flavor violation may have implications for the pattern of slepton masses
and mixings. Possible signals for this mixing in the decays of sleptons
produced at the LHC are discussed. The sensitivity expected is compared
to that of rare decays such as $\tau\to \mu\gamma$.

\end{abstract}

\section{Introduction\label{sec:intro}}

Recent results from Super Kamiokande\cite{Kajita:1999bw}  and 
earlier measurements from Soudan-2\cite{Allison:1997yb} and
IMB\cite{Casper:1991ac} on the relative fluxes of electron and
muon type neutrinos produced by the interaction of cosmic rays 
in the atmosphere show a deficit of muon-type neutrinos. These 
results are consistent with the oscillation of $\nu_\mu$ into
$\nu_\tau$ provided that  the two mass eigenstates are
fully mixed,  $\sin^22\theta \approx 1$, and their mass
difference is of order $\Delta m^2 \sim 10^{-3}
\hbox{eV}^2$. 
Models that can accommodate this mixing have
lepton flavor violation (LFV) and other lepton number violating
processes which are likely to be observable, such as $\tau \to \mu
\gamma $ and $\tau \to \mu \mu^+\mu^-$. 
It has been suggested\cite{Ellis:1999uq, Feng:2000wt, Hisano:1996cp} 
that  supersymmetric models can naturally accommodate the flavor
violation.

If there is mixing of neutrinos, then one would in
general expect mixing of sleptons and sneutrinos in SUSY.
In general, SUSY models can have
significant flavor mixing. The slepton mass matrix
is a $6\times 6$ matrix involving the three flavors of left
and right
sleptons and has the form
$$
\tilde{\ell}^*_{M\, i}
(M^2_{\tilde{\ell}})^{MN}_{ij} \tilde{\ell}_{N\, j} =
\matrix{( \tilde{\ell}^*_{L\, i} & \tilde{\ell}^*_{R\, k})  }
\left(\matrix{M^2_{L,\, ij} & M^2_{LR,\,il} \cr
M^2_{LR,\,{jk}}  & M^2_{R,\,kl}} \right)
\left(\matrix{\tilde{\ell}_{L\, j} \cr \tilde{\ell}_{R\, l}} \right)
\ ,
$$
where $M,N = L,R$ label chirality and $i,j,k,l = 1,2,3$ are
generational indices.  $M^2_L$ and $M^2_R$ are the supersymmetry
breaking, chirality conserving, slepton masses. The chirality breaking 
terms, $M^2_{LR,\,jk}$, contain a flavor diagonal piece $m_l \mu \tan\beta$,
where $m_l$ 
is the corresponding charged lepton mass, $\mu$ is the supersymmetry
conserving Higgs mass parameter and $\tan\beta$ is the usual
ratio of Higgs vacuum expectation values. Additional supersymmetry
breaking terms ($A$ terms) also appear in the chirality mixing term. These 
cannot be flavor diagonal if, 
as we require, the neutrino mass matrix is not diagonal in the flavor
basis. However, these terms are expected to be proportional to charged 
lepton masses. We will therefore make the simplifying assumption that
these terms are relevant only for the third generation. In order to
accommodate the $\nu_\mu$ into
$\nu_\tau$ mixing, there must be significant mixing in the
stau, smuon sector\cite{Ellis:1999uq}. 

In the MSSM (minimal supersymmetric standard model) there is no lepton
number violation. It can be incorporated by adding right handed
neutrinos ($N_i$) with couplings to the left handed lepton doublets
$L_j$ of the form $\lambda_{ij}N_iL_jH$ where $H$ is a Higgs
doublet. Then radiative corrections at one loop will induce lepton
number violating terms in the mass matrices for the left
sleptons. There can also be similar soft supersymmetry breaking $A$
terms. We therefore assume the following form for
the slepton mass matrix in the $\tilde e_L, \tilde\mu_L, \tilde\tau_L,
\tilde e_R, \tilde\mu_R, \tilde\tau_R$ basis \cite{Hisano:1996cp}:
$$ 
M_{\tell\tell}^2 = \left[{
\begin{array}{cccccc}
M_{L}^2+D_{L} & 0 & 0 & 0 & 0 & 0 \\
0 & M_{L}^2+D_{L} & M_{\mu\tau}^2 & 0 & 0 & 0 \\
0 & M_{\mu\tau}^2 & M_{\tau_L}^2+D_{L} & 0 & 0 & m_\tau\bar{A}_\tau \\
0 & 0 & 0 & M_{R}^2+D_{R} & 0 & 0 \\
0 & 0 & 0 & 0 & M_{R}^2+D_{R} & 0 \\
0 & 0 & m_\tau\bar{A}_\tau & 0 & 0 & M_{\tau_R}^2+D_{R}
\end{array}
}\right]
$$
where $D_{L}= -\frac12(2M_W^2-M_Z^2)\cos2\beta$,
$D_{R}=(M_W^2-M_Z^2)\cos2\beta$, and
$\bar{A}_\tau=A_\tau-\mu\tan\beta$. Minimal supersymmetric models such
as  SUGRA are recovered by neglecting $M_{\mu\tau}^2$. 
The Super Kamiokande results suggest
$$
\delta \equiv {M_{\mu\tau}^2 \over M_{L}^2} = {\cal O}(1)\,.
$$

It has been suggested that this large flavor violation in the slepton
sector could be observed indirectly via the decay $\tau\to
\mu\gamma$\cite{Ellis:1999uq, Feng:2000wt}.  By contrast, this paper
considers direct signals from the observation of lepton flavor
non-conservation in the SUSY sector at the LHC. We will demonstrate that, 
in certain cases,
the direct observation is more sensitive than the flavor violating tau
decay. Section~\ref{sec:model} defines a sample point based on the
minimal SUGRA model that has been studied in detail, and
Section~\ref{sec:tau} explains how sensitivity to $\tau-\mu$ flavor
violation can be observed at the LHC for this model.
Section~\ref{sec:rare} compares the sensitivity for this direct search
with that from the rare decay $\tau \to \mu\gamma$.
Section~\ref{sec:general} presents results indicating the range of
parameters in the minimal SUGRA model for which such direct searches
might be feasible. Finally, Section~\ref{sec:concl} presents a summary
and conclusions.

\section{Example -- Model Parameters\label{sec:model}}

We consider a minimal SUGRA\cite{SUGRA,SUGRArev} point with $m_0=100\,\GeV$,
$\mhalf=300\,\GeV$, $A_0=300\,\GeV$, $\tan\beta=10$, and $\sgn\mu=+$. We
use the ISAJET\cite{ISAJET} implementation of the SUGRA model. This point
is 
identical to one of those studied in detail\cite{Hinchliffe:1997iu,luc}
(SUGRA Point 5)  except for the value of $\tan\beta$. The
larger $\tan\beta$ gives a light Higgs mass of $113\,\GeV$, consistent
with the current LEP limit\cite{LEPhiggs}. 
The masses are calculated with ISAJET~7.49,
which assumes $\delta=0$. A modification was made to make available  the
complete slepton mass matrix, giving access to all the soft mass
parameters at the electroweak scale {\it viz.}:
\begin{eqnarray*}
&M_L=236.0\,\GeV,\quad M_R=153.3\,\GeV,\quad \mu=360.2\,\GeV&\\
&M_{\tau_L}=235.2\,\GeV,\quad M_{\tau_R}=150.7\,\GeV,\quad A_\tau=94.9\,\GeV&
\end{eqnarray*}
The mass eigenvalues for $\delta=0$ are
$$
m_{\tell_L}=231.76,\quad m_{\tell_R}=156.26,\quad 
m_{\ttau_1}=145.36,\quad m_{\ttau_2}=232.96\,\GeV.
$$
($\ttau_1$ is the lighter of the two mass eigenstates resulting from the
mixing of $\ttau_L$ and $\ttau_R$.)  Note that in  SUGRA type models the
$\tell_R$ are generally lighter than the $\tell_L$.  The dominant final
states for $\tchi_2^0$ decay are $\lsp Z^0$, $\tilde e_R e$ and
$\tilde\mu_R \mu$, and $\ttau_1\tau$ which have branching ratios of
6.9\%, 12.8\%  (for both $e$ and $\mu$), and 66.1\% respectively. The
$\tchi_1^\pm$ branching ratios are 45.3\% for $\lsp W$ and 54.4\% for
$\ttau_1\nu_\tau$.

Direct Drell-Yan production of sleptons at the LHC has a small
cross section (270 fb for all sleptons in the case considered)
and is difficult to separate from SUSY backgrounds. In contrast the
total SUSY rate, made up largely of squark and gluino production is
~100 times larger. Gluinos (of mass 706 GeV) decay to squarks (average
mass 620 GeV) with an almost 100\%
branching ratio and left-squarks have a 30\% branching ratio to
$\tchi_2^0$ (of mass 212 GeV). Hence the
most copious source of observed dileptons from 
sleptons in this case is $\tchi_2^0 \to 
\tell\ell$ where the $\tchi_2^0$ arises in
squark decay. This will always be the case provided that the decay is
allowed. If only $\tell_R$ is produced in this way (in the case
studied  $\tell_L$ is heavier than $\tchi_2^0$), 
then the main effect of $\delta\ne0$ is to add a component of
$\tilde\mu_L$ in $\ttau_1$ so that $\tchi_2^0 \to \ttau_1 \mu$ and
$\ttau_1 \to \mu \lsp$.  The $\lsp$ mass is 116 GeV. This decay 
gives a $\mu^\pm\tau^\mp$ signal and
produces
an asymmetry between $\mu^\pm\tau^\mp$ and $e^\pm\tau^\mp$ final states.

\begin{figure}[t]
\dofig{3in}{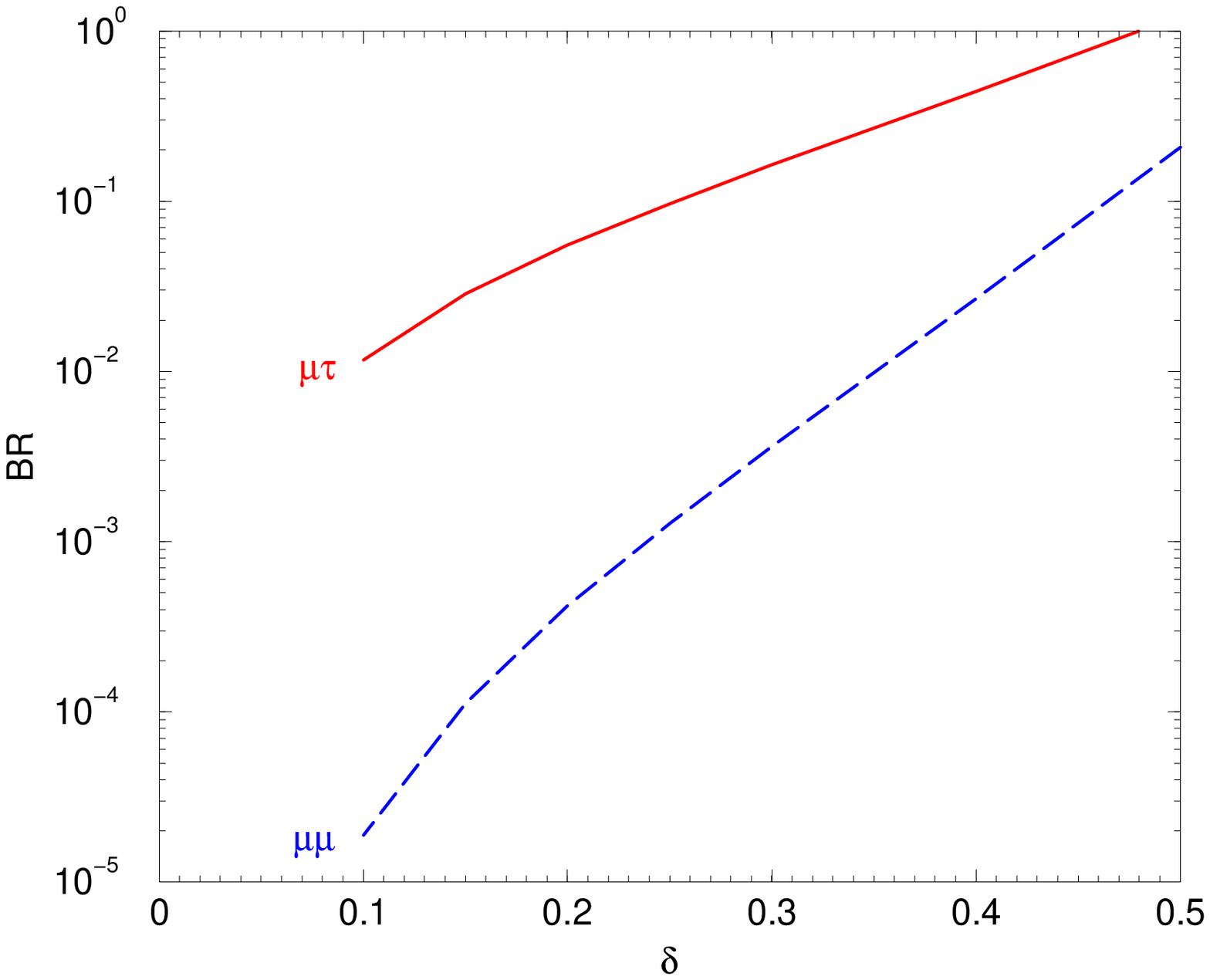}
\caption{Branching ratios for $\tchi_2^0 \to \lsp \mu\tau$ (solid) and for
$\tchi_2^0 \to \lsp \mu\mu$ through an intermediate $\ttau_1$ (dashed) 
as a function of the mixing parameter $\delta$.
\label{c10br}}
\end{figure}

The slepton mass matrix was rediagonalized using the soft mass parameters
given above plus the mixing term $\delta$, inducing mixings among
slepton flavors and hence flavor-violating decays such as $\tchi_2^0 \to
\ttau \mu$ and $\ttau \to \lsp \mu$. The resulting  branching ratios for
$\tchi_2^0 \to \lsp \mu\tau$ and for $\tchi_2^0 \to \lsp \mu\mu$ through
an intermediate $\ttau_1$ with at least one lepton-flavor violating decay are
shown in Figure~\ref{c10br}.  The latter decay is uninteresting both
because it is small and because it cannot be separated
from the $\tchi_2^0\to\ttau_1\tau\to\tau\tau\lsp$ final state for
which both taus 
decay to muons. We therefore concentrate on the
$\mu\tau$ decay followed by a hadronic tau decay. Note that the mass
shifts in the lightest states are  very small. For $\delta=0.1$, the masses
of the lightest states change by $\sim 100$ MeV; the 
heavier states are more strongly affected, their masses become 
221 and 243 GeV; the mass of $\tilde{e}_L$ is, of course,  unchanged at
 231 GeV.

A sample of 600000 SUSY events was generated with
ISAJET~7.49, equivalent to an integrated
luminosity of $\sim 25 $ fb$^{-1}$. 
The detector response is parameterized by Gaussian resolutions
characteristic of the ATLAS\cite{ATLAS} detector without any 
non-Gaussian tails.
  In the central region of
rapidity we take separate resolutions for the electromagnetic (EMCAL)
and hadronic (HCAL) calorimeters, while the forward region uses a
common resolution:
\begin{eqnarray*}
{\rm EMCAL} &\quad& 10\%/\sqrt{E} \oplus 1\%, |\eta|\, <3 \,\nonumber\\
{\rm HCAL}  &\quad& 50\%/\sqrt{E} \oplus 3\%, |\eta|\, < 3\,\nonumber\\
{\rm FCAL}  &\quad& 100\%/\sqrt{E} \oplus 7\%, |\eta|\, > 3\,\nonumber
\end{eqnarray*} 
All energy and momenta are measured in GeV. A uniform segmentation 
$\Delta\eta = \Delta\phi = 0.1$ is used with no
transverse or longitudinal 
shower spreading. Both ATLAS\cite{ATLAS} and CMS\cite{CMS} have
finer segmentation over most of the rapidity range.  An oversimplified
parameterization of the muon momentum resolution of the ATLAS detector
including  both the inner tracker and the muon system measurements is
used, {\it viz}
$$
\delta p_T/p_T = \sqrt{0.016^2+(0.0011p_T)^2}
$$
In the case of electrons we take a momentum resolution obtained by
combining the electromagnetic calorimeter resolution above with a
tracking resolution of the form
$$
\delta p_T/p_T= \left(1+{0.4\over(3-\abs{\eta})^3}\right)
\sqrt{(0.0004p_T)^2+0.0001}
$$
This provides a slight improvement over the calorimeter alone.
Missing transverse energy is calculated by taking the magnitude of the
vector sum of the transverse energy deposited in the calorimeter
cells. 
Jets are found using GETJET\cite{ISAJET} with a simple
fixed-cone algorithm. The jet multiplicity in SUSY events is rather
large, so we have  used a cone size of 
$$
R = \sqrt{(\Delta\eta)^2+(\Delta\phi)^2} = 0.4
$$ 
unless otherwise stated. Jets are required to have at least $p_T >
20$~GeV; more stringent cuts are often used.  All leptons are required
to be isolated and have some minimum $p_T$ and $\abs{\eta}< 2.5$.  The
isolation requirement is that no more than 10~GeV of additional $E_T$
be present in a cone of radius $R = 0.2$ around the lepton to reject
leptons from $b$-jets and $c$-jets.

\section{Final States Involving Taus\label{sec:tau}}

SUSY production at the LHC is dominated by the production of squarks and
gluinos, which decay via cascades to the lightest SUSY particle $\lsp$,
which escapes the detector. Events were therefore selected by requiring
multiple jets plus large $\etmiss$:
\begin{itemize}
\item   $\ge4$ jets with $p_{T,1}>100\,\GeV$ and $p_{T,2,3,4}>50\,\GeV$;
\item   $\Meff \equiv \etmiss + p_{T,1} + p_{T,2} + p_{T,3} + p_{T,4} 
>800\,\GeV$;
\item   $\etmiss > 0.2\Meff$;
\end{itemize}
The same analysis was applied to the Standard Model background.
ISAJET~7.44 was used to generate 250000 events each of
$t\overline{t}$, $Wj$, $Zj$, and
$WW$ and 2500000 QCD jets, divided among five $p_T$ bins covering $50 <
p_T < 2400\,\GeV$. The cut on $\Meff$ was chosen to make this background
fairly small but still visible. These background samples represent much
smaller integrated luminosity that the signal sample, so the
resulting statistical fluctuations are large when they are scaled to
the appropriate integrated luminosity.

\begin{figure}[t]
\dofig{3in}{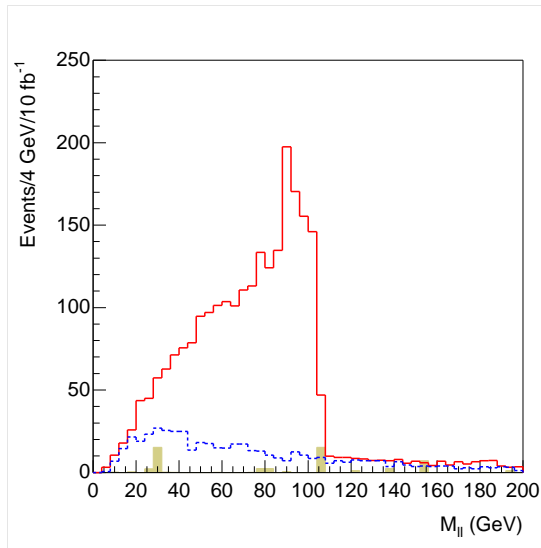}
\caption{Dilepton mass distribution: OS (opposite sign) SF (same
  flavor) signal (solid), OSOF signal
(dashed), and Standard Model OSSF (shaded). \label{c10mll}}
\end{figure}

\begin{figure}[t]
\dofig{3in}{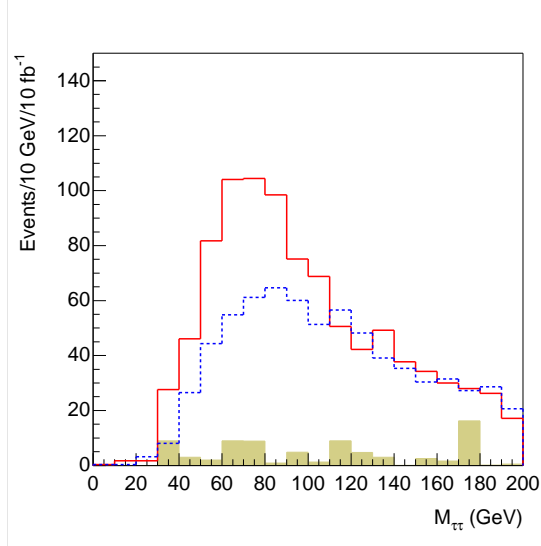}
\caption{Visible $\tau_h\tau_h$ mass for $\tau_h^+\tau_h^-$ signal (solid),
$\tau_h^\pm\tau_h^\pm$ signal (dashed), and Standard Model $\tau_h^+\tau_h^-$
(shaded). Only hadronic tau decays are included. \label{c10mtautau}}
\end{figure}

\begin{figure}[t]
\dofig{3in}{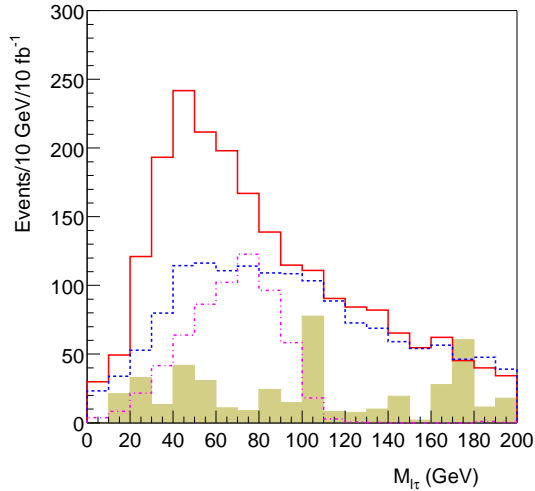}
\caption{Visible mass distributions for $\ell^\pm\tau_h^\mp$ signal
(solid), $\ell^\pm\tau_h^\pm$ signal (dashed), $\mu^\pm\tau_h^\mp$ from LFV
decays with $BR=10\%$ (dot-dashed), and Standard Model $\ell^\pm\tau_h^\mp$
(shaded). Only hadronic tau decays are included.\label{c10mltaulfv10}}
\end{figure}

For events that pass this selection, an additional requirement 
that there be two isolated leptons with $p_T>10 $ GeV and
$\abs{\eta} < 2.5$ is made.
The dilepton mass distribution for opposite sign, same flavor pairs is 
shown in Figure~\ref{c10mll} and shows the
characteristic endpoint for slepton mediated decay. The end-point
occurs at
$$
M_{\ell\ell}^{\rm max} = \sqrt{(M_{\tchi_2^0}^2 - M_{\tell}^2)
(M_{\tell}^2 - M_{\lsp}^2) \over M_{\tell}^2} = 95.1\,\GeV\,.
$$
A small $Z^0$ peak is also visible, primarily from the 6.9\% branching
ratio for $\tchi_2^0 \to \lsp Z^0$. The events in the opposite sign,
opposite flavor sample
are due to decays of the type $\tchi_2^0 \to 
\ttau_1\tau \to \tau^+\tau^- \lsp\to 
e^+\mu^-\nu_e\overline{\nu}_\mu\nu_\tau\overline{\nu}_\tau$ and from
events with leptons arising from $\tchi_1^{\pm}$ decay which are not
subject to the kinematic constraint.

Selection of hadronic $\tau$ decays was based on the actual
hadronic tau decays  plus the reconstruction efficiency and jet
rejection based on the full simulation analysis\cite{ianhtau} 
done for an earlier study\cite{Hinchliffe:2000zc}. Pile-up effects have
been ignored, so the results are applicable only at low luminosity,
$\sim 10^{33}\,\cmsec$.
For each hadronic $\tau$ decay with $p_T>p_{T,{\rm min}}=20\,\GeV$
and $\eta<2.5$, the direction was found. A matching jet with $\eta<2.5$
was then sought with $\Delta R<0.4$ and $p_{T,\tau}>0.8p_{T,{\rm jet}}$.
If such a jet was found, it was tagged as a $\tau$ with a probability of
66\%; its charge was assigned correctly with a probability of 92\%. Jets
not so tagged were assigned as $\tau$'s using an approximate 
parameterization of the
jet/$\tau$ rejection shown in Figure~9-31 of Ref.~\citenum{TDR},
$$
R = \left(0.971p_T^{\frac32} - 49\,\GeV\right)^{\frac53(1-\epsilon_\tau)}
$$
where $\epsilon_\tau$ is the hadronic $\tau$ efficiency and $p_T$ is in
GeV. We choose $\epsilon_\tau=66\%$. The efficiency and the results
using the parameterization of\cite{ianhtau} are not very different.
Note that we are actually using  
the hadronic jet resolution for all $\tau$ decays here, not the improved
resolution for a subset of decays as in the earlier analysis
\cite{Hinchliffe:2000zc} which was focussed on the invariant mass of a 
pair of taus, both decaying hadronically. Since this
analysis relies mainly on counting events, the resolution on the 
measurement of the hadronic  $\tau$  decay products is not 
crucial.

\begin{figure}[t]
\dofig{3in}{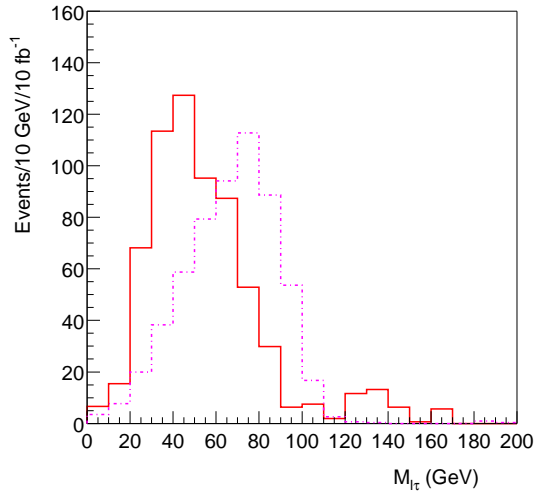}
\caption{Visible mass distributions for $\ell^\pm\tau_h^\mp -
\ell^\pm\tau_h^\pm$ (solid) and $\mu^\pm\tau_h^\mp$ from LFV decays with
$BR=10\%$ (dot-dashed). All the Standard Model backgrounds considered
here should cancel in this plot up to statistical fluctuations.
\label{c10mltaudiflfv}}
\end{figure}

The visible (hadronic) $\tau\tau$ mass for hadronic $\tau$ decays
(hereafter denoted by $\tau_h$) is shown in Figure~\ref{c10mtautau}.
Here events were required to have two $\tau_h$
candidates in addition to the cuts listed above. Note that one or both 
of these candidates could be the same jets that were used in the
initial event selection. The background from misidentified QCD jets
is approximately random in sign and so cancels in the $\tau^+\tau^- -
\tau^\pm\tau^\pm$ combination. To the extent that $\tg\tg$ and $\tg\tq$
production dominates, so does the background where both $\tau$'s come from
$\tchi_1^\pm$ decay. Figure~\ref{c10mtautau} shows an excess of
$\tau^+\tau^-$ pairs ending approximately at the endpoint of the
dilepton distribution (from $\tchi_2^0\to \ttau\tau$) 
shown in Figure~\ref{c10mll}. The structure is
less clear due to the energy carried off by the neutrinos.

Even for $\delta=0$ there is a substantial rate for real
$\ell^\pm\tau_h^\mp$ pairs from $\tau$ pairs produced by the decay chain
$\tchi_2^0\to \ttau_1 \tau\to \lsp \tau^+\tau^-$. Two independent
chargino decays can also give $\ell\tau$ pairs of either sign.
Misidentified QCD jets give additional $\ell\tau_h$ pairs; the signs of
$\tau\tau$ pairs from such misidentified jets should be random, and this
is assumed here. The LFV mixing from $\delta\ne0$ produces additional
$\mu\tau_h$ pairs, 92\% of which have signs that are correctly identified.
Figure~\ref{c10mltaulfv10} shows the OS and SS
backgrounds and the signal for an assumed 10\% branching ratio for the
direct decay $\tchi_2^0 \to \lsp \tau^\pm \mu^\mp$. This corresponds to
$\delta= 0.25$ (see Figure~\ref{c10br}).  Since ISAJET assumes lepton
flavor conservation, the decay of interest was simulated by finding
events with two $\tau$'s consistent with $\tchi_2^0 \to \lsp \tau\tau$
and with at least one hadronic $\tau$ decay and then replacing the other
$\tau$ with a muon with a probability equal to the assumed branching
ratio. This is an excellent approximation since, as is indicated above,
the mass shift due to mixing is very small.  Figure~\ref{c10mltaudiflfv}
shows the signal  and the sign-subtracted background.
The latter should cancel in the subtracted $\mu\tau-e\tau$ distribution
up to statistical fluctuations; the LFV signal occurs only in the
$\mu\tau$ channel. The distribution from the LFV signal
has a peak at larger values of invariant mass than that of the lepton
flavor conserving process  as the 
lepton  in the latter must arise from the decay of a tau and is
therefore softer than that the muon from $\tchi_2^0\to \mu\tau\lsp$

In Figure~\ref{c10mltaulfv10} for $50 < M_{\ell\tau} <
100\,\GeV$, there are 1089 OS and 707 SS events, with equal numbers of
$e\tau$ and $\mu\tau$, and there are 518 lepton flavor violating
$\mu\tau$ events, 92\% of which are correctly identified as to sign.
Hence in this mass range we expect
\begin{eqnarray*}
N(\mu^\pm\tau^\mp) &=& =0.92(.5\times1089 + 518) = 977 \\
N(e^\pm\tau^\mp) &=& .5\times1089\times 0.92 = 501
\end{eqnarray*}
Adding the errors in quadrature, we would then measure
$$
N(\mu^\pm\tau^\mp) - N(e^\pm\tau^\mp)= 476 \pm 39
$$
giving a $12.2\sigma$ excess for $10\,\fbi$.
The statistical  $5\sigma$ limit for
$30\,\fbi$ would be a branching ratio of $2.3\%$,
corresponding to $\delta\approx0.1$. The signal to background ratio is better
in the sign subtracted distributions, but the statistical errors are
larger. 
A determination of the branching ratio  from the observation of an
excess of $\mu\tau$ events  requires a detailed understanding of the
systematic uncertainties. The scaling of these results to the design
LHC luminosity also requires a careful study of the effects of pile up 
of low-$p_T$ hadronic events.
 It is interesting to remark that the sensitivity that we
obtain from direct slepton decay is comparable to that claimed for a
lepton collider operating at 500 GeV center of mass energy for
$30\,\fbi$\cite{Hisano:1999fd}.

\section{Comparison with Rare Decays\label{sec:rare}}

The decay $\tau \to \mu\gamma$ is sensitive to the same lepton flavor
violation as the signal considered here. The approximate formula of
Ref.~\citenum{Feng:2000wt},
$$
BR(\tau\to\mu\gamma) \approx 1.1\times10^{-6}
\left({\delta\over1.4}\right)^2
\left({100\,\GeV\over M_\tell}\right)^4
$$
gives for $\delta=0.1$ and $M_\tell=150\,\GeV$
$$
BR(\tau\to\mu\gamma) \approx 1\times10^{-9}\,.
$$
However this result is sensitive to the details of the mass spectra and
mixings; cancellations can occur resulting in much smaller
rates\cite{Hisano:1996cp,Hisano:1997qq}.  For comparison, the current
bound is $1.1\times10^{-6}$\cite{pdg}. The total production of $W
\to\tau\nu$ at the LHC is about $10^{9}$ events for an integrated luminosity of
$100\,\fbi$. A study\cite{stry} carried out for the ATLAS detector
concluded that a 90\% confidence level upper limit of
$0.6\times10^{-6}$
on $
BR(\tau\to\mu\gamma)$ could be reached using $W\to \tau\nu$ events for an integrated luminosity of $30\,\fbi$ corresponding
to three years of running at low luminosity. The process is background
limited from QED final state radiation in the decays  $W \to \mu\nu$ and $W \to
\tau\nu \to \mu\nu\nu$.  These contribute approximately 50 events for
$30\,\fbi$.

\section{Generalization of Results\label{sec:general}}

We have demonstrated, by using a specific example, how flavor
violation in the slepton sector can be observed at the LHC. It would
be difficult
to repeat the analysis for a large number of
other cases. In order to estimate the generality of the method we have
adopted a simpler approach. The crucial feature is the production of
staus in decays of $\tchi_2^0$. For fixed values of $A$, $\tan\beta$
and $\mu$, the parameter space of $m_0$ and $m_{1/2}$ was scanned. At
each point, all masses and branching ratios were computed assuming
that there is no slepton flavor mixing. If the gluino is lighter than
the up and down squarks, then the product branching ratio $BR(\tilde g
\to \chi_2^0 +X)\times BR(\tchi_2^0 \to \tilde \tau_1 \tau)$ is
computed, if the gluino is heavier than
the up and down squarks the combination $(BR(\tilde u_L
\to \chi_2^0 +X)+BR(\tilde d_L
\to \chi_2^0 +X))\times BR(\tchi_2 \to \tilde \tau_1 \tau)/2$ is
calculated.
The total supersymmetric cross section is determined at a few points
in the parameter space and can be approximated by
$$
\sigma=1.79\times10^{13}\left(0.1m_0+m_{1/2}\right)^{-4.8}\,{\rm pb}\,.
$$
This formula agrees with the cross-section to within 25\% for all
relavant paramters; the dependence on $A$, $\tan\beta$ and $\mu$ is
small. An approximation of the total production rate for
$\tilde{\tau_1}$ from $\tchi_2^0$ arising 
from squark and gluino production and decay can then
be obtained from the product of the cross-section and the combined
branching ratio. 

\begin{figure}[t]
\dofigs{3in}{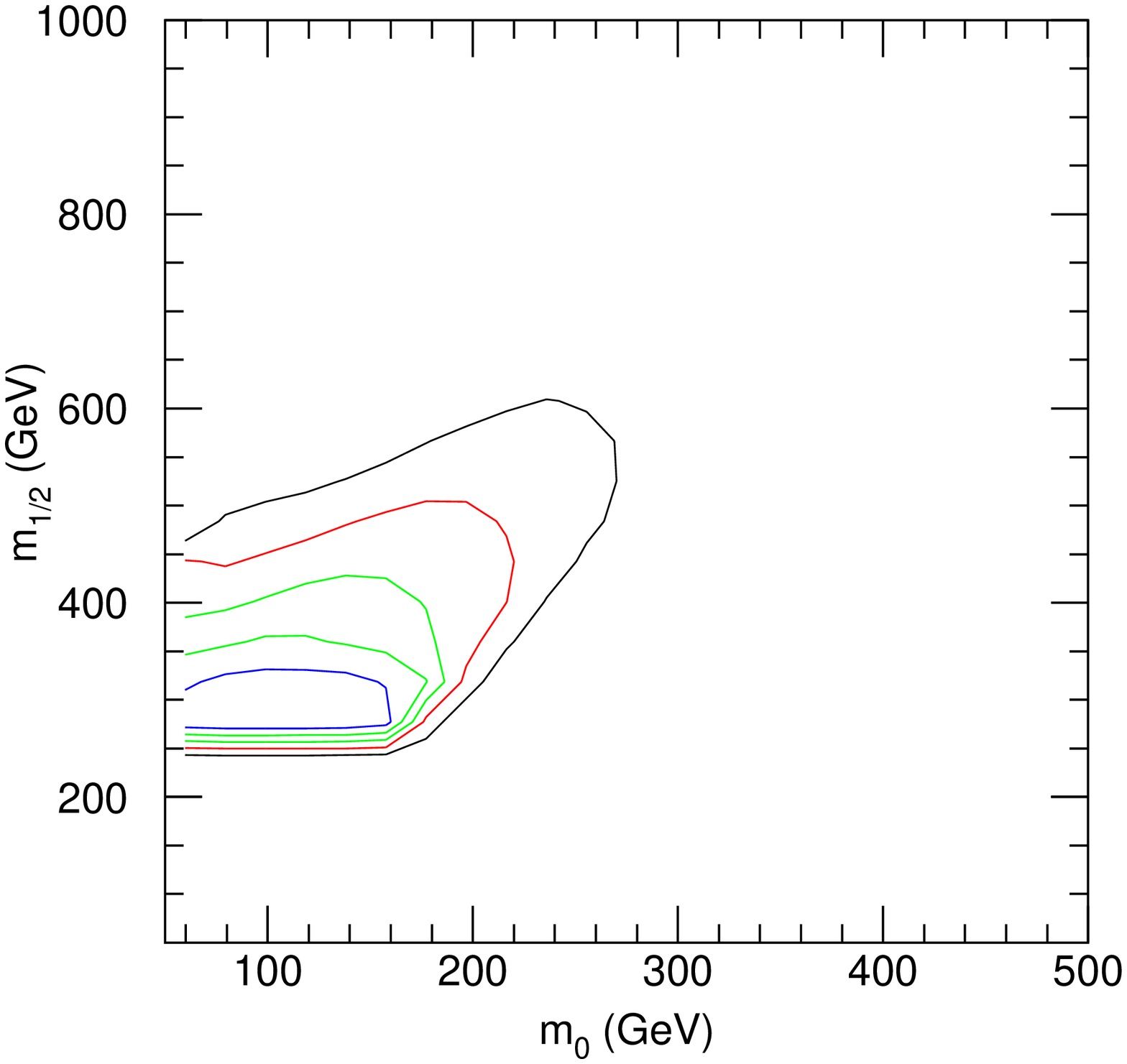}{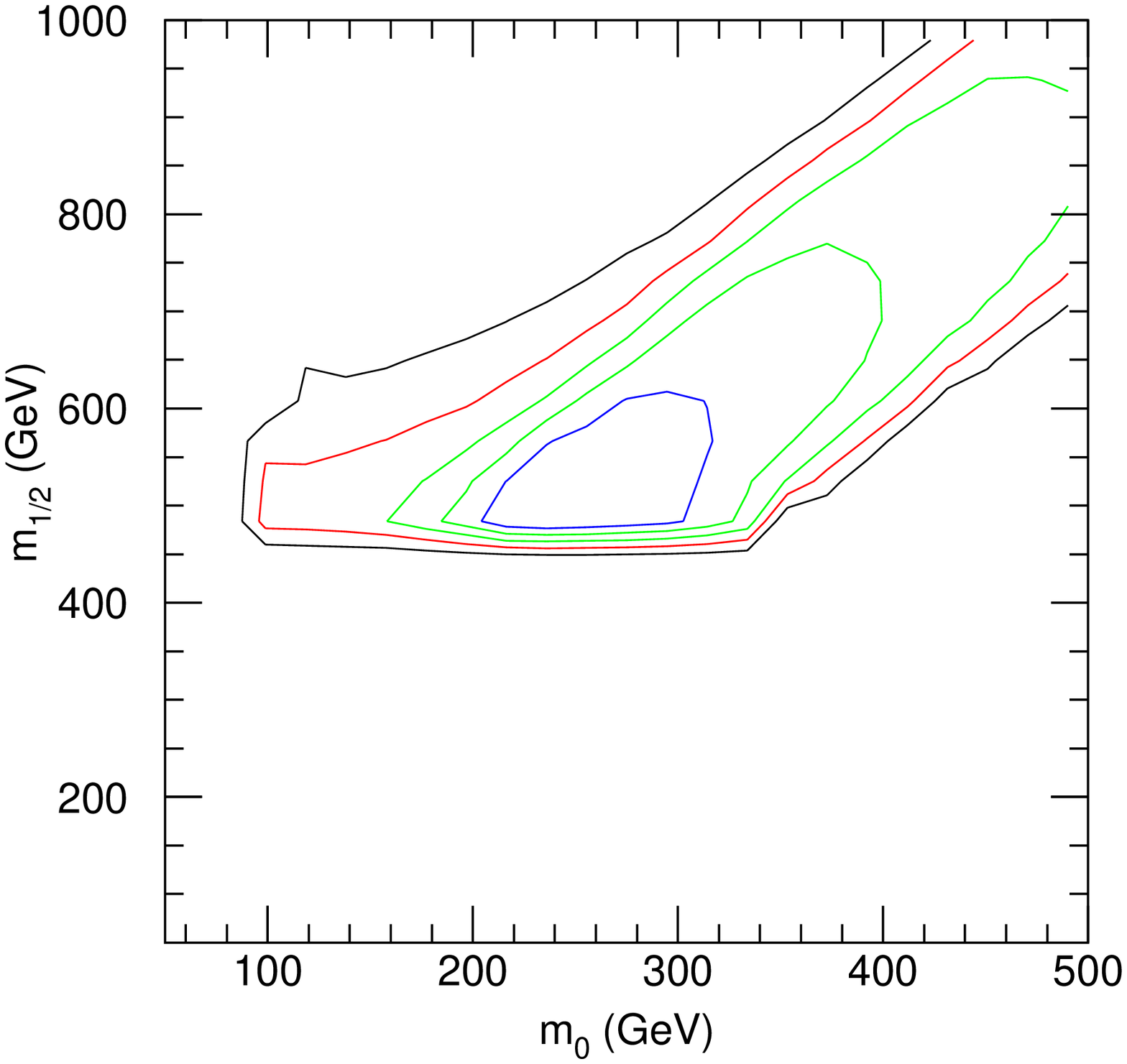}
\caption{Contours showing an estimate of the total production rate for
$\tilde{\tau_1}$ as a function of $m_0$ and $m_{1/2}$ for $A=0$,
$\tan\beta=3$, and $\sgn\mu=+$ (left) and $\sgn\mu=-$ (right). The contours
(outer to inner) are at 0.03 0.1 0.3, 1 and 3 pb; the scale is
logarithmic. \label{c3rate}}
\end{figure}

\begin{figure}[t]
\dofigs{3in}{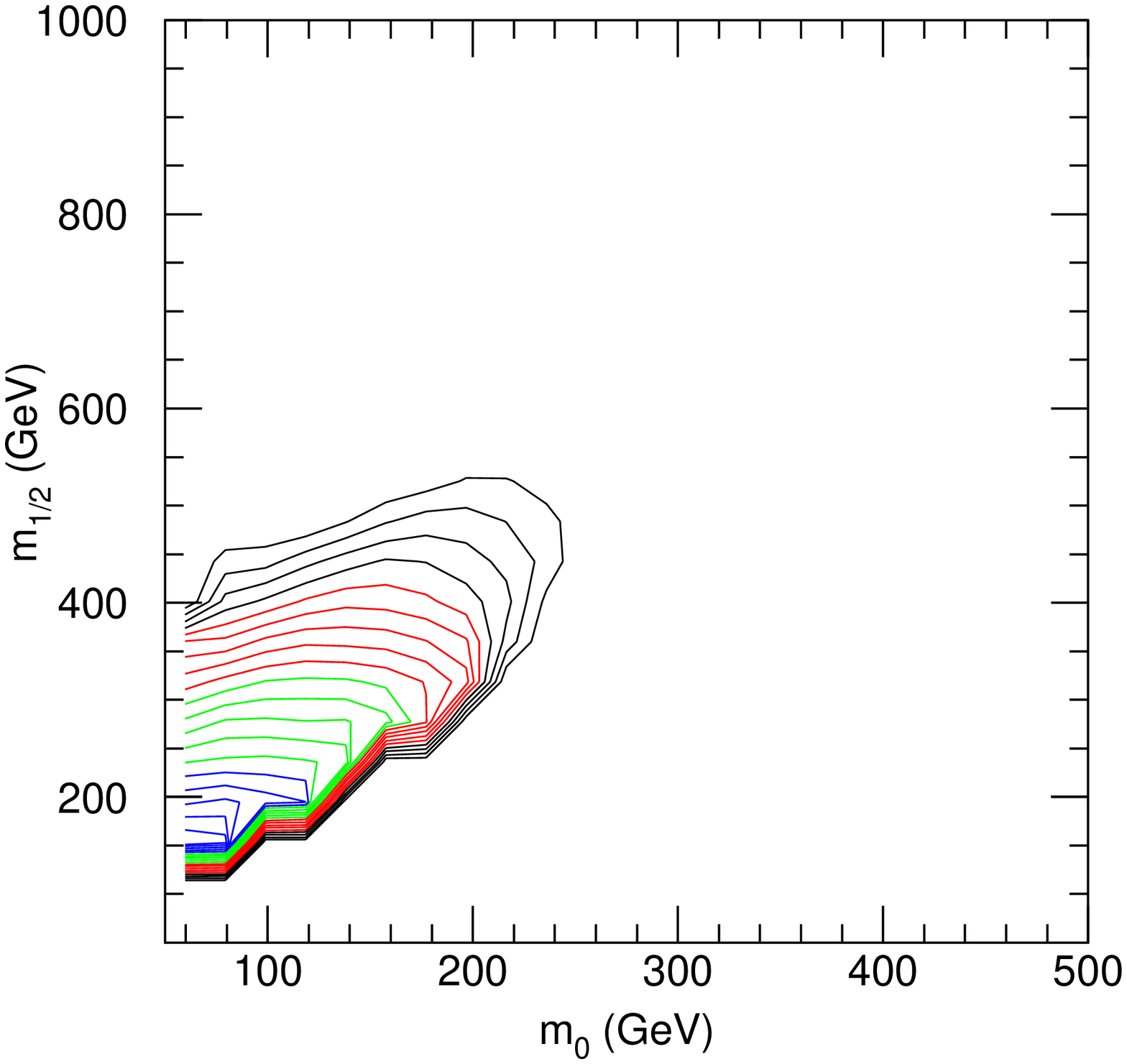}{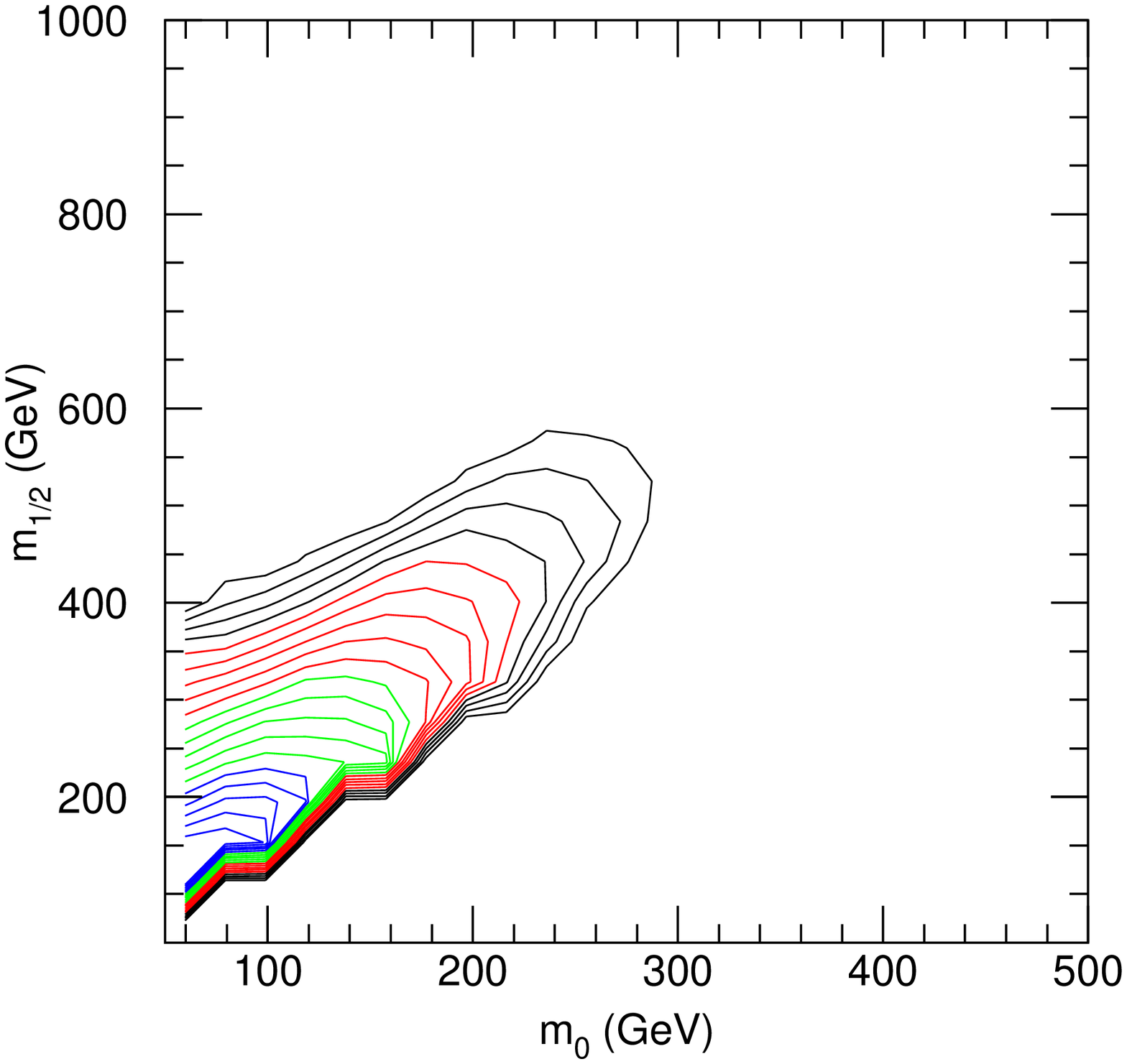}
\caption{Contours showing an estimate of the total production rate for
$\tilde{\tau_1}$ as a function of $m_0$ and $m_{1/2}$ for $A=0$,
$\tan\beta=10$, and $\sgn\mu=+$ (left) and $\sgn\mu=-$ (right). The
contours (outer to inner) extend from  0.1 pb to 100 pb and are
uniformly spaced on a logarithmic scale. \label{c10rate}}
\end{figure}

\begin{figure}[t]
\dofigs{3in}{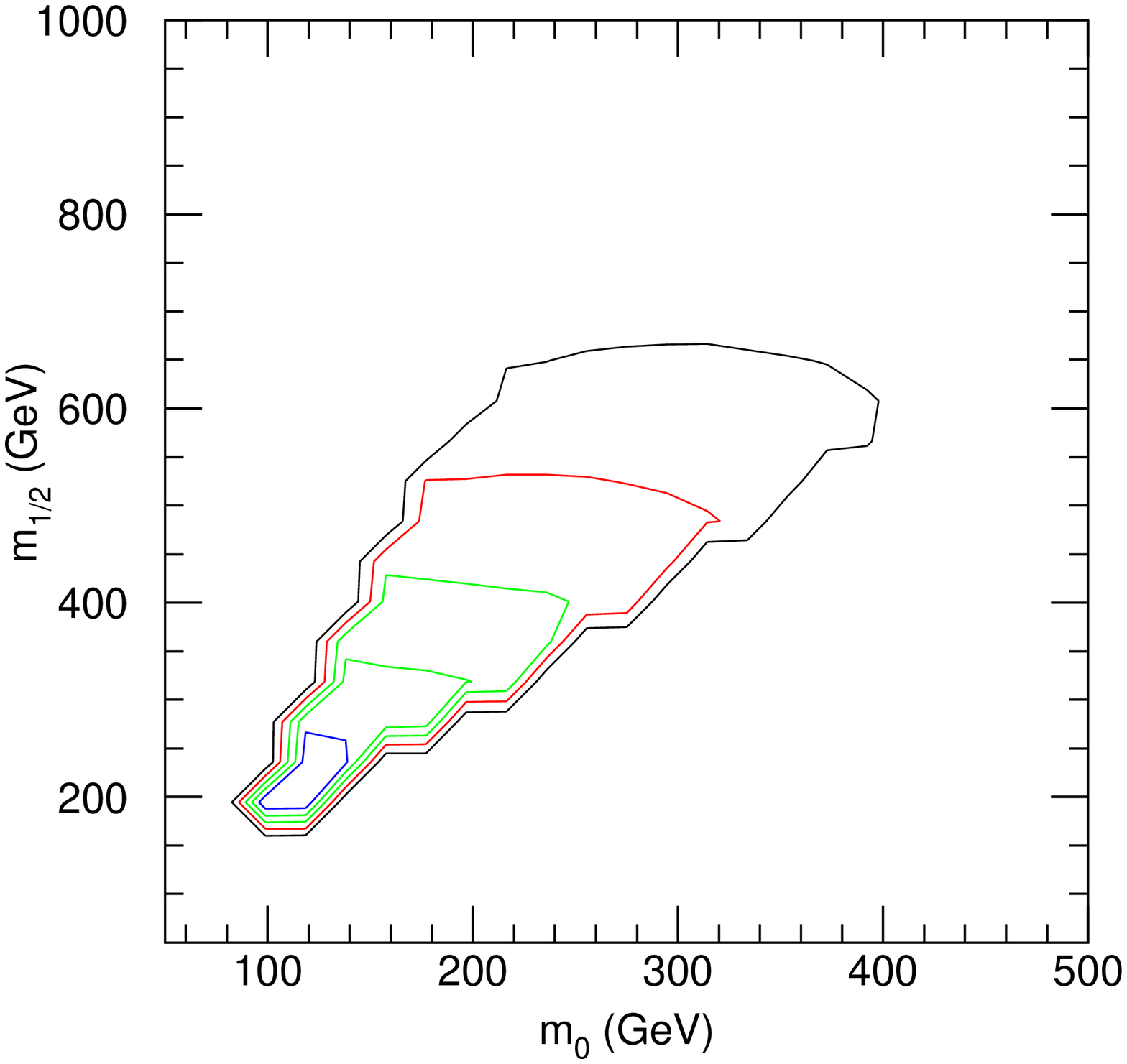}{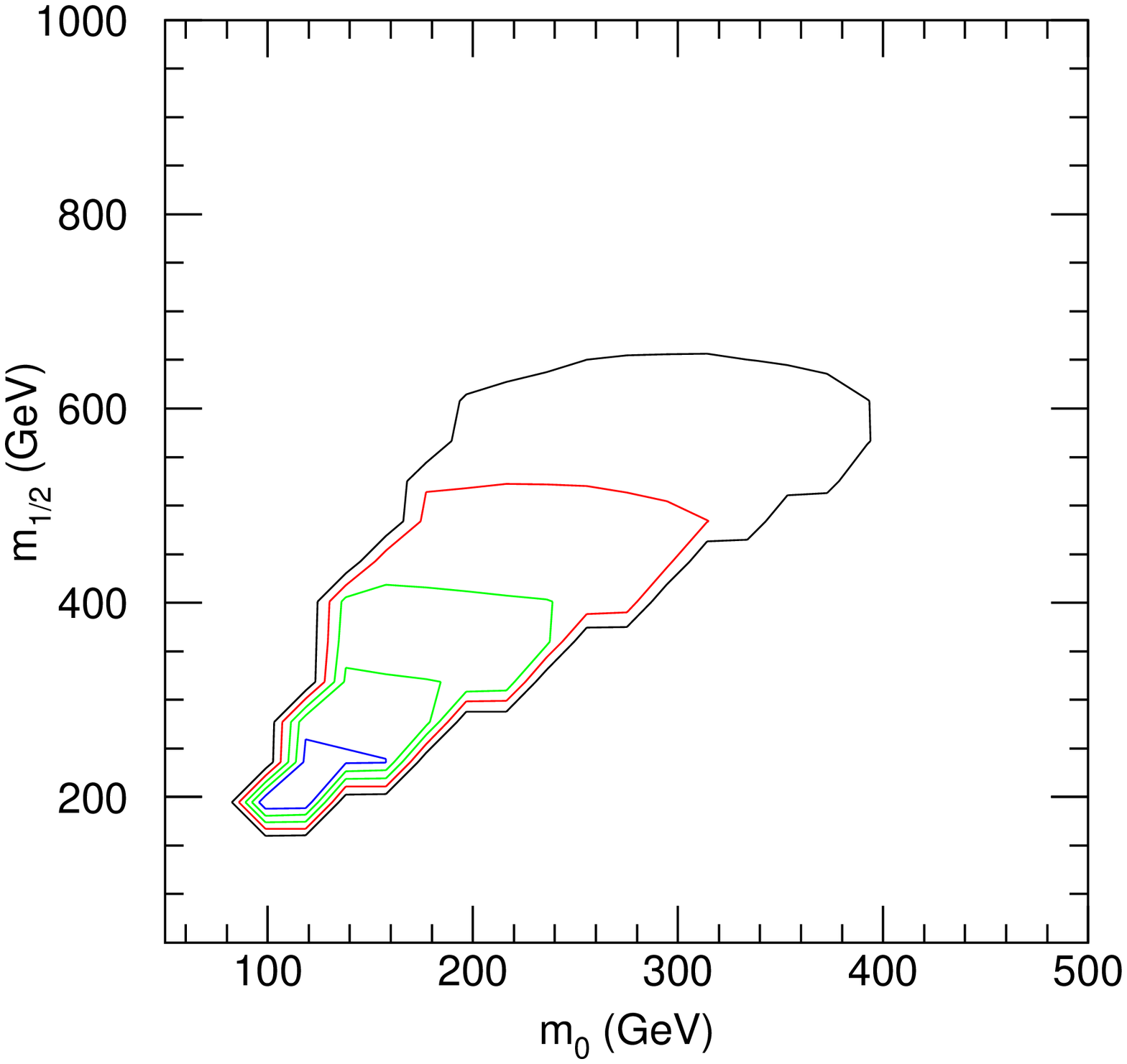}
\caption{Contours showing an estimate of the total production rate for
$\tilde{\tau_1}$ as a function of $m_0$ and $m_{1/2}$ for $A=0$,
$\tan\beta=30$, and $\sgn\mu=+$ and $\sgn=-$ (right). The contours
(outer to inner) are at 0.1 0.3, 1, 3 and 10 pb; the scale is logarithmic.
\label{c30rate}}
\end{figure}

\begin{figure}[t]
\dofigs{3in}{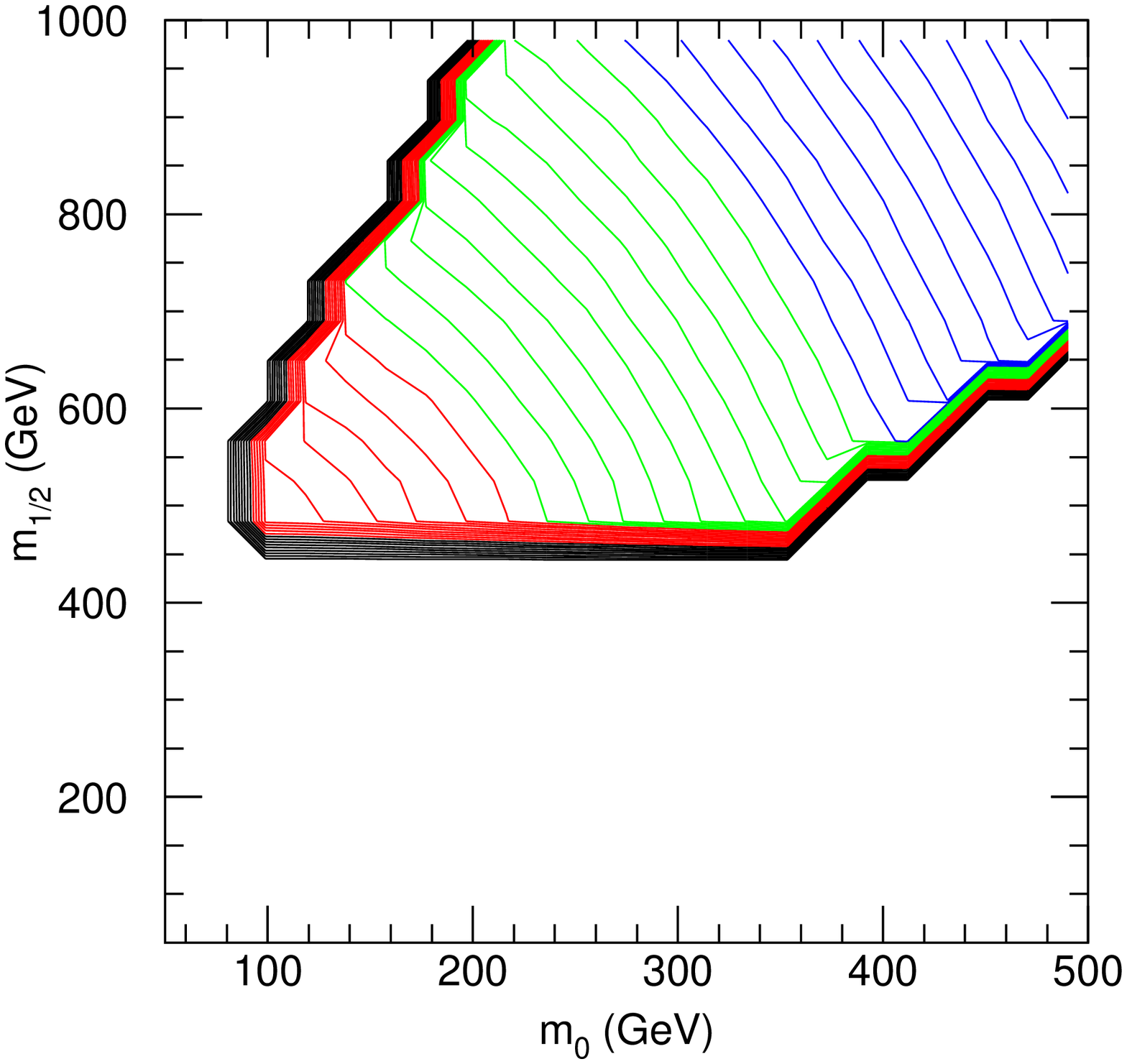}{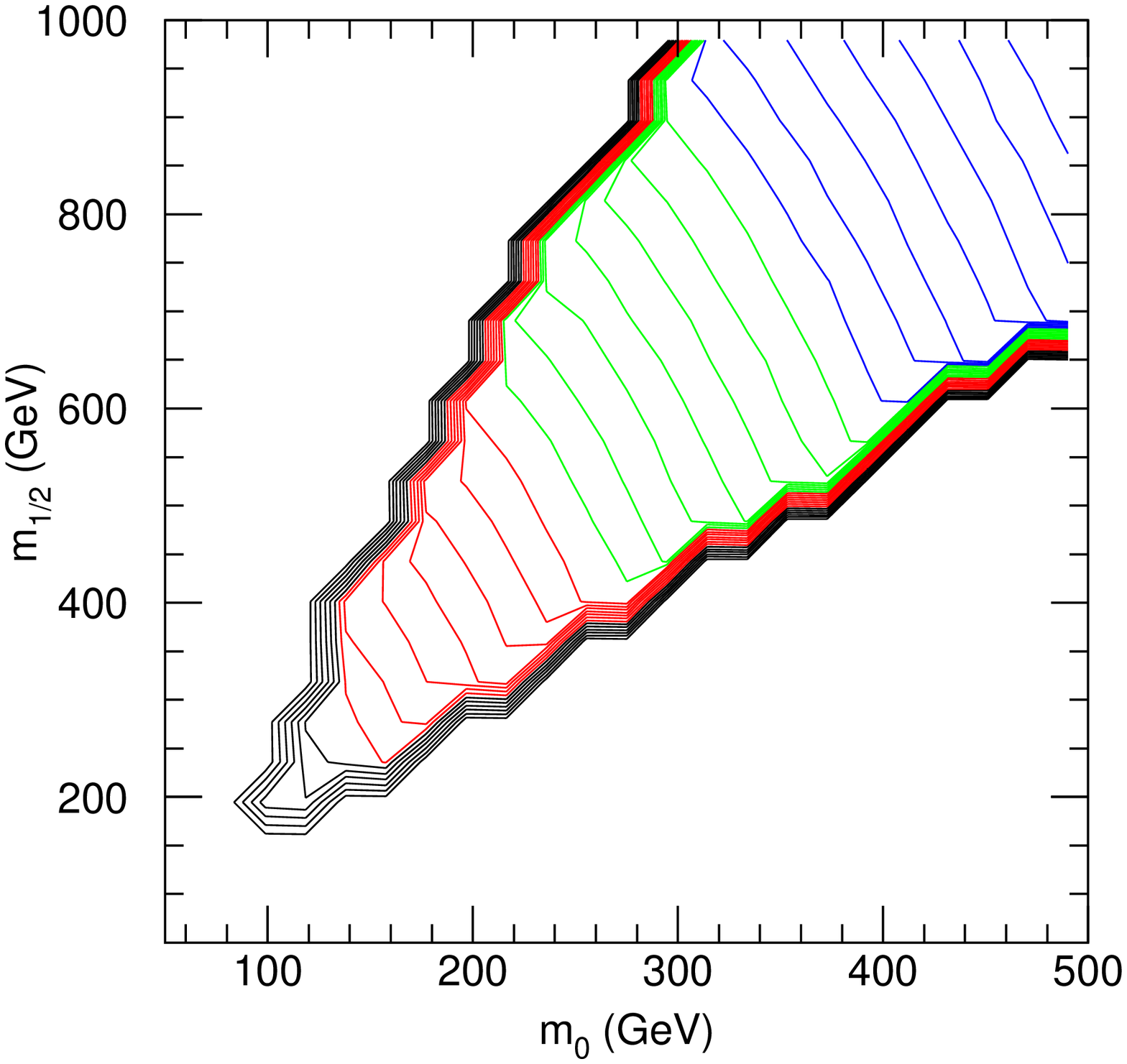}
\caption{Contours showing the mass of $\tilde{\tau_1}$ as a function of
$m_0$ and $m_{1/2}$ for $A=0$ and $\sgn\mu=-$. Left: $\tan\beta=3$;
the contours are 20 GeV apart, from 130~GeV in the bottom corner to
610~GeV in the top corner. Right: $\tan\beta=30$;
the contours are 25 GeV apart, from 75~GeV in the bottom corner to
530~GeV in the top corner. The lower bound in $m_{1/2}$ in the left
figure results from the requirement $M_h>103\,\GeV$; this bound is less
stringent for $\sgn\mu=+$. \label{tmass}}
\end{figure}

Figures~\ref{c3rate} through \ref{c30rate} show the resulting
rates, and Figure~\ref{tmass} shows two sample sets of contours of the 
$\tilde{\tau_1}$ mass. The light Higgs mass is required
to be larger than 103~GeV in making these plots; this accounts for the
cut off at smaller values of $m_{1/2}$ for  $\tan\beta =3$.  Rasing
the limit to 107 GeV excludes the region below $m_{1/2}=365$ GeV 
for $\sgn\mu=+$. The
current bound
on the Higgs mass for such a small value of $\tan\beta$ is
approximately 112 GeV \cite{LEPhiggs}. At this value   $m_{1/2}$ is so
large that the production rate is less than 0.1 fb. The effect of this
bound should be taken {\it cum grano salis} as uncomputed higher order
corrections could modify the relationship between the model parameters
and the Higgs mass.
The Higgs limit has no impact
on the  plots with $\tan\beta>5$.
The lack of rate at large values of $m_{1/2}$ is due to the large
gluino and squark masses and the consequent small production rate. The
region where $m_0\simge m_{1/2}$ is not accessible as there the decay
$\tchi_2^0 \to \tau\tilde{\tau}$ is not allowed kinematically.
For comparison, the case studied in detail would correspond to
a $\ttau_1$ rate of $~\sim 5$ fb according to Figure~\ref{c10rate}. A
comparison of this value with the rates shown in
Figure~\ref{c10mtautau} shows that the combination of acceptance and
tau detection efficiency is approximately 2\%.  This is not expected
to vary dramatically over the relavent parameter space as the
kinematics of production and decay are broadly similar. 
The acceptance will be less near the boundary where
$m_{\tchi_2^0}\sim m_{\tilde{\tau}}$ as the resulting taus will be
soft. The acceptance is greater at larger values of $m_0$ or $m_{1/2}$
where  events are more central and the resulting taus more
energetic. At a
luminosity of $10^{33}$ cm$^{-2}$ sec$^{-1}$ a meaningful measurement
should  be possible if the rate is larger than 1 pb. An examination of
the plots shows that this covers a sizable fraction of parameter
space. A detailed quantitative assessment of the sensitive region is
not straightforward. This is because the background arises from
supersymmetry itself and not from standard model sources. The
background therefore varies across parameter space, so a more detailed
sensitivity estimate would require repeating the full analysis at each
point.

The identification of $\tau_h$ decays requires tight cuts on the shower
shape and track multiplicity, so it becomes more difficult at high
luminosity, $10^{34}\,\cmsec$. For high SUSY masses, however, the
$\tau_h$ typically have higher $p_T$ and so are more easily
distinguished from QCD jets. Extending the analysis described here to
high luminosity requires a much more detailed and detector-specific
study of pileup effects than has been done here. If it proves to be
possible, then the parameter space where the rate is larger than 0.1 pb
should be accessible. This covers almost all of the physically
interesting region where the decay $\tchi_2^0\to \tau\tilde{\tau_1}$ is
allowed.

\section{Conclusion\label{sec:concl}}

We have demonstrated that the observation of lepton flavor
violation  may be possible at the LHC. We have focussed on the
case where the largest violation occurs in the $\mu-\tau$
sector. While this is motivated by the atmospheric neutrino problem
this violation is  harder to detect than that occuring in the  $\mu-e$
sector since the
detection of hadronically decaying taus is involved. If sleptons are
produced in the decays of squarks and gluinos the rates are large
enough that the resulting sensitivity is greater than can be reached
in rare decays of the type $\tau\to\mu\gamma$. This corresponds to the
fraction of parameter space where $m_{1/2}>m_0$. In the SUGRA model
the fraction of parameter space where these decays occur with a rate
large enough for observation is large. 

If sleptons are not
produced in these decays, the sensitivity will be much less. A study
of the possibility of observing direct slepton production for the mass 
spectrum considered (without flavor violation)\cite{luc}. Stringent
jet veto cuts are required to reduce the backgrounds from other SUSY
events. Approximately 50 events from $\ell_l$ production were found in 
$30\,\fbi$ and a mass sensitivity of 20 GeV obtained. Note that this
sensitivity is comparable to the mass shifts induced in the  $\tilde{\ell_l}$
sector for $\delta=0.1$. A more detailed study is needed before a
definite conclusion can be made, but it appears that this process
might also be more sensitive than the $\tau \to \mu \gamma$ decay.

The importance of $\tau_h$ decays for SUSY extends beyond the lepton
flavor violating signal considered here. If $\tan\beta\gg1$, then the
decays $\tchi_2^0 \to \ttau_1\tau$ and $\tchi_1^\pm \to \ttau_1^\pm \nu$
can be the only allowed two-body decays and so have branching ratios
close to unity; then $\tau$ signatures dominate\cite{TDR}. For any value
of $\tan\beta$, $\tau_h$ signatures can provide information both on the
splitting between $\ttau_1$ and $\tell_R$ and on the mixing between
$\ttau_R$ and $\ttau_L$\cite{hptau}. It is therefore important to continue
a detailed study of $\tau_h$ signatures for the LHC detectors.

\bigskip

We are grateful to Georges Azuelos, John Ellis, Fabiola Gianotti,
 Ryszard Stroynowski
and Takeo Moroi
for useful discussions.  This work was supported in part by the Director, Office of Science,
Office of Basic Energy Research, Division of High Energy
Physics of the U.S. Department of Energy under Contracts
DE--AC03--76SF00098 and DE-AC02-98CH10886.  Accordingly, the U.S.
Government retains a nonexclusive, royalty-free license to publish or
reproduce the published form of this contribution, or allow others to
do so, for U.S. Government purposes.

\end{document}